# A gravitationally unstable gas disk of a starburst galaxy 12 billion years ago

K. Tadaki[1], D. Iono[1,2], M. S. Yun[3], I. Aretxaga[4], B. Hatsukade[5], D. Hughes[4], S. Ikarashi[6], T. Izumi[1], R. Kawabe[1,2,7], K. Kohno[5,8], M. Lee[9], Y. Matsuda[1,2], K. Nakanishi[1,2], T. Saito[9], Y. Tamura[9], J. Ueda[1], H. Umehata[11], G. W. Wilson[3], T. Michiyama[1,2], M. Ando[1,2] and P. Kamieneski[3]

[1] *National Astronomical Observatory of Japan, 2-21-1 Osawa, Mitaka, Tokyo 181-8588, Japan*
[2] *SOKENDAI (The Graduate University for Advanced Studies), 2-21-1 Osawa, Mitaka, Tokyo 181-8588, Japan*
[3] *University of Massachusetts, Department of Astronomy, 710 North Pleasant Street, Amherst, MA 01003, USA*
[4] *Instituto Nacional de Astrofisica, Opticay Electronica (INAOE), Luis Enrique Erro 1, Sta. Ma. Tonantzintla, Puebla, Mexico*
[5] *Institute of Astronomy, The University of Tokyo, 2-21-1 Osawa, Mitaka, Tokyo 181-0015, Japan*
[6] *Kapteyn Astronomical Institute, University of Groningen, P.O. Box 800, 9700AV Groningen, The Netherlands7*
[7] *Department of Astronomy, The University of Tokyo, 7-3-1 Hongo, Bunkyo-ku, Tokyo 133-0033, Japan*
[8] *Research Center for the Early Universe, The University of Tokyo, 7-3-1 Hongo, Bunkyo, Tokyo 113-0033, Japan*
[9] *Division of Particle and Astrophysical Science, Nagoya University, Furocho, Chikusa, Nagoya 464-8602, Japan*
[10] *Max-Planck-Institute for Astronomy, Königstuhl 17, D-69117, Heidelberg, Germany*
[11] *The Institute of Physical and Chemical Research (RIKEN), 2-1 Hirosawa, Wako-shi, Saitama 351-0198, Japan*

**Submillimeter bright galaxies in the early Universe are vigorously forming stars at ~1000 times higher rate than the Milky Way. A large fraction of stars is formed in the central 1 kiloparsec region[1-3], that is comparable in size to massive, quiescent galaxies found at the peak of the cosmic star formation history[4], and eventually the core of giant elliptical galaxies in the present-day Universe. However, the physical and kinematic properties inside a compact starburst core are poorly understood because dissecting it requires angular resolution even higher than the Hubble Space Telescope can offer. Here we report 550 parsec-resolution observations of gas and dust in the brightest unlensed submillimeter galaxy at z=4.3. We map out for the first time the spatial and kinematic structure of molecular gas inside the heavily dust-obscured core. The gas distribution is clumpy while the underlying disk is rotation-supported. Exploiting the high-quality map of molecular gas mass surface density, we find a strong evidence that the starburst disk is gravitationally unstable, implying that the self-gravity of gas overcomes the differential rotation and the internal pressure by stellar radiation feedback. The observed molecular gas would be consumed by star formation in a timescale of 100 million years, that is comparable to those in merging starburst galaxies[5]. Our results suggest that the most extreme starburst in the early Universe originates from efficient star formation due to a gravitational instability in the central 2 kpc region.**

In the past two decades since the first discovery of submillimeter bright galaxies (SMGs) at high redshift[6-7], their global physical properties, such as redshift, gas mass, and kinematics, have extensively been studied to understand the origin of the extreme starburst[8-12]. For one of the brightest, and unlensed submillimeter galaxies COSMOS-AzTEC-1 at z=4.3, we obtained the highest-resolution data cube of CO $J=4-3$ emission line using the Atacama Large Millimeter/submillimeter Array (ALMA). AzTEC-1 is intensively forming stars at a rate of $1169^{+11}_{-274}$ $M_\odot$ yr$^{-1}$ and has a compact starburst with a half-light radius of $R_{1/2}=1.1\pm0.1$ kpc in the 860 μm continuum[13]. We spatially resolve the CO line at 0.08″-resolution (550 pc in the physical scale) to study the morphology and the kinematics of molecular gas within the central 2 kpc region. Fig. 1 shows ALMA maps of CO line and dust continuum fluxes at 3.2 mm and 860 μm, velocity field, and velocity dispersion in AzTEC-1. The spatial distributions of the CO and the 3.2 mm continuum emission independently confirm the existence of two off-center clumps--clump-2 and clump-3--, reported in the 860 μm continuum observations[13]. Several 0.15-0.3″ resolution observations found that SMGs and optically-selected massive galaxies are associated with a very compact dusty star-forming region with $R_{1/2}$=1-2 kpc[1-3,14]. The higher-resolution view, however, demonstrates that the central structure of molecular gas and dust is more complicated than just a single, compact component. Such molecular gas clumps are also seen in a central disk of a gravitationally-lensed star-forming galaxy at z=3, SDP 81[15,16].

We furthermore make a 0.06″-resolution CO cube and a 0.05″-resolution 860 μm continuum map with different visibility weightings, which are more sensitive to clump structures than the underlying disk component (see Methods). The velocity-integrated CO maps show that the molecular gas clumps are aligned with the dusty star-forming clumps in the 860 μm continuum (Fig 2). They are the second and third brightest clumps among 11 clumps previously identified at 860 μm[13]. As the brightest clump is very close to the nuclei, it is hard to isolate this even at 0.06″-resolution. Other faint star-forming clumps are not detected in the CO data, probably due to the lack of the sensitivity.

We fit the CO spectra of the clumps with a single Gaussian component to measure line widths of FWHM=250±50 km s$^{-1}$ for clump-2 and FWHM=240±50 km s$^{-1}$ for clump-3, which are one or two orders larger than those of giant molecular clouds (GMC) in nearby galaxies[17]. The CO flux is $S_{co}dv$=0.056±0.009 Jy km s$^{-1}$ for clump-2 and $S_{co}dv$=0.042±0.007 Jy km s$^{-1}$ for clump-3, indicating that each clump contains only a few percent of the total gas mass. Assuming that the CO-to-H$_2$ conversion factor of $\alpha_{CO}$=0.8 $M_\odot$ (K km s$^{-1}$ pc$^2$)$^{-1}$ and the CO excitation of $R_{41}$=0.91 are homogeneous over the galaxy, we obtain gas masses of $M_{CO,gas}$=(2.2±0.3)×10$^9$ $M_\odot$ and $M_{CO,gas}$=(1.7±0.3)×10$^9$ $M_\odot$ (see Methods), which are 3-5 orders larger than the virial mass of GMCs. These giant clumps are totally different from GMCs in nearby galaxies.

We fit the CO cube with dynamical models to derive the kinematic properties in the central region. The observed velocity field of the underlying component is well characterized by a rotating disk with a half-light radius of $R_{1/2}$=1.05$^{+0.02}_{-0.02}$ kpc, a deprojected maximum rotation velocity of $v_{max}$=227$^{+5}_{-6}$ km s$^{-1}$, and a local velocity dispersion of $\sigma_0$=74$^{+1}_{-1}$ km s$^{-1}$. The starburst disk is rotation-dominated with a deprojected rotation velocity-to-velocity dispersion ratio of $v_{max}/\sigma_0$=3.1±0.1. In the local Universe, 80% of massive early-type galaxies with a stellar mass of log($M_{star}/M_\odot$)>11.8 show dispersion-dominated stellar kinematics with $v_{max}/\sigma_0$<1 while less massive ones are rotation-dominated[18,19]. Given the large stellar mass of $M_{star}$=(9.7$^{+0.0}_{-2.5}$)×10$^{10}$ $M_\odot$ (see Methods), AzTEC-1 probably resides in the most massive system at z=4 and eventually evolve into the most massive early-type galaxies at z=0. If molecular gas and stars share the same kinematics, the observed rotating disk suggests that the most massive galaxies do not lose a large fraction of the angular momentum in the formation phase and does it by the subsequent evolution such as major mergers[20].

Until recently, clumpy rotating disks at high-redshift were discovered by observations of ionized gas[21]. Now, we can do the same thing with higher-resolution observations of molecular gas using ALMA. Both observational and numerical studies show that giant clumps are formed via gravitational instability in outskirt of gas-rich disks and migrate inward due to the dynamical friction[22,23]. Using the ALMA maps of CO flux and velocity dispersion without correction for beam smearing (Fig. 1), we compute the local Toomre $Q$-parameter, expressing a balance between self-gravity of molecular gas and turbulent pressure by stellar radiation and other sources. A thick, rotating disk can become unstable against local axisymmetric perturbations for $Q<Q_{cri}$=0.7[24]. The measured local $Q$ values are much below $Q_{cri}$ over the entire disk, indicating that gas collapses through the gravitational instability in the interclump regions. On the other hand, Q<$Q_{cri}$ at the clump location means that gas is bounded rather than being gravitationally unstable. We also derive radially-averaged Q parameters using the best-fit kinematic parameters with correction for beam smearing and inclination. Here, the uncertainties of the derived Q parameter mainly come from gas mass measurements. We approach this issue with three independent observations of CI (2-1), CO (4-3) line, and dust continuum. All three methods indicate that the Q value is less than unity in the central 2.5 kpc region, even with taking into account some variations in carbon abundance, CO excitation, and gas-to-dust mass ratio (Fig 3).

In the current framework of galaxy evolution, galaxies self-regulate star formation with a marginally unstable disk[25,26]. If a galaxy disk are unstable with $Q<Q_{cri}$, intense stellar radiation would temporarily boost a turbulent pressure and then heats up the disk until $Q>Q_{cri}$. Once the disk is stable, star formation would become inefficient, leading to a drop in turbulent pressure. At the same time, additional gas accretion may increase a molecular gas mass surface density in the disk. When the increased self-gravity of gas overcomes the decreased pressure, the disk becomes unstable. Thus, galaxies keep a marginally unstable disk with $Q\sim Q_{cri}$. However, in AzTEC-1, the stellar radiation pressure is unlikely to support the self-gravity of gas, resulting in small $Q$ values over the entire disk. The local velocity dispersion hardly increases with a rise in SFR surface density (Fig 3), suggesting that stellar feedback by intense star formation does not control velocity dispersion in molecular gas. We also find that the velocity dispersion of $\sigma\sim 100$ km s$^{-1}$ in the two clumps is not extremely high, compared to the rest of the disk. Our results imply that star-forming clumps are bound and are not disrupted by radiative feedback. On the other hand, there is a strong correlation between molecular gas mass surface density and SFR surface density, fitted by a linear function of $\log(\Sigma_{SFR}/M_\odot$ yr$^{-1}$kpc$^{-2})=(1.4\pm0.2)\times\log(\Sigma_{gas}/M_\odot$pc$^{-2})+(-3.6\pm0.7)$. The gas mass surface density is extremely high with $\log(\Sigma_{gas}/M_\odot$pc$^{-2})=3.8$-$4.4$, that is similar to nearby starburst galaxies[5]. The implied gravitational instability comes as a consequence of the strong concentration of molecular gas.

In such a gravitationally unstable gas disk, molecular clouds are expected to be efficiently converted into stars. The gas depletion timescale, defined as gas mass divided by star format rate, in the starburst disk is comparable to the galaxy-averaged timescales in nearby starburst galaxies (Fig. 3). The molecular gas reservoir will be consumed by star formation within 100 million years, that is by a factor of ~10 shorter than gas depletion time scales in typical star-forming galaxies at z=1-3[27] and comparable to those in nearby merging galaxies like Arp220 and Arp299[5]. A nucleus in a gravitationally-lensed star-forming galaxy also has a very short depletion timescale[16]. The extreme starburst at high-redshift would occur in a very short time, resulting in episodic bright lights in submillimeter wavelength. Otherwise, it requires gas fueling into the central region to maintain the current active star formation.

It is still uncertain how a large amount of molecular gas is strongly concentrated in the central 2 kpc region. A gas-rich major merger is the most straightforward scenario as several numerical simulations successfully reproduce the physical properties, including the compact gas distribution and the enhanced star forming activity, of SMGs[28]. A rotating disk does not necessarily reject the major merger scenario since nearby merger remnants mostly show a clear rotation[29]. In addition to a past gas-rich major merger, multiple gas-rich minor mergers or clumpy gas stream are likely to contribute to a gas transport from outside the galaxy to the central 2 kpc region. On the other hand, we do not have a direct evidence for a major merger in AzTEC-1. For isolated galaxies, it requires a non-axisymmetric structure such as a spiral arm and a bar to lose the angular momentum of gas and transport a large amount of gas into a galaxy center. AzTEC-1 does not have such a non-axisymmetric structure. To know the roles of major mergers in extreme starburst, we need to investigate morphological and kinematic structures in a large sample of high-redshift SMGs using high-resolution (<0.1″) and sensitive observations with ALMA.

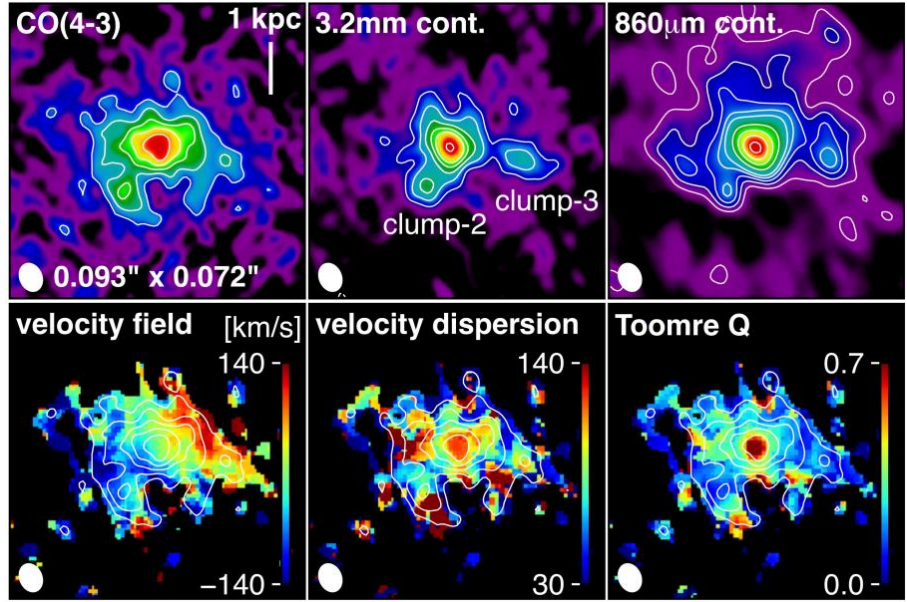

**Figure 1 | CO morphology and kinematics for AzTEC-1.** ALMA maps of CO(4-3) line, 3.2 mm, and 860 μm continuum emission, velocity field, velocity dispersion and Toomre Q parameter. The numbers in bracket refer to the rest-frame wavelength. The angular resolution is all 0.093″×0.072″. The CO flux is integrated in the velocity range of -315 km s$^{-1}$<$v$<+315 km s$^{-1}$. All contours are plotted at every 2σ from 3σ to 11σ and at every 5σ from 11σ. In the bottom three panels, the CO flux distribution is overplotted as white contours.

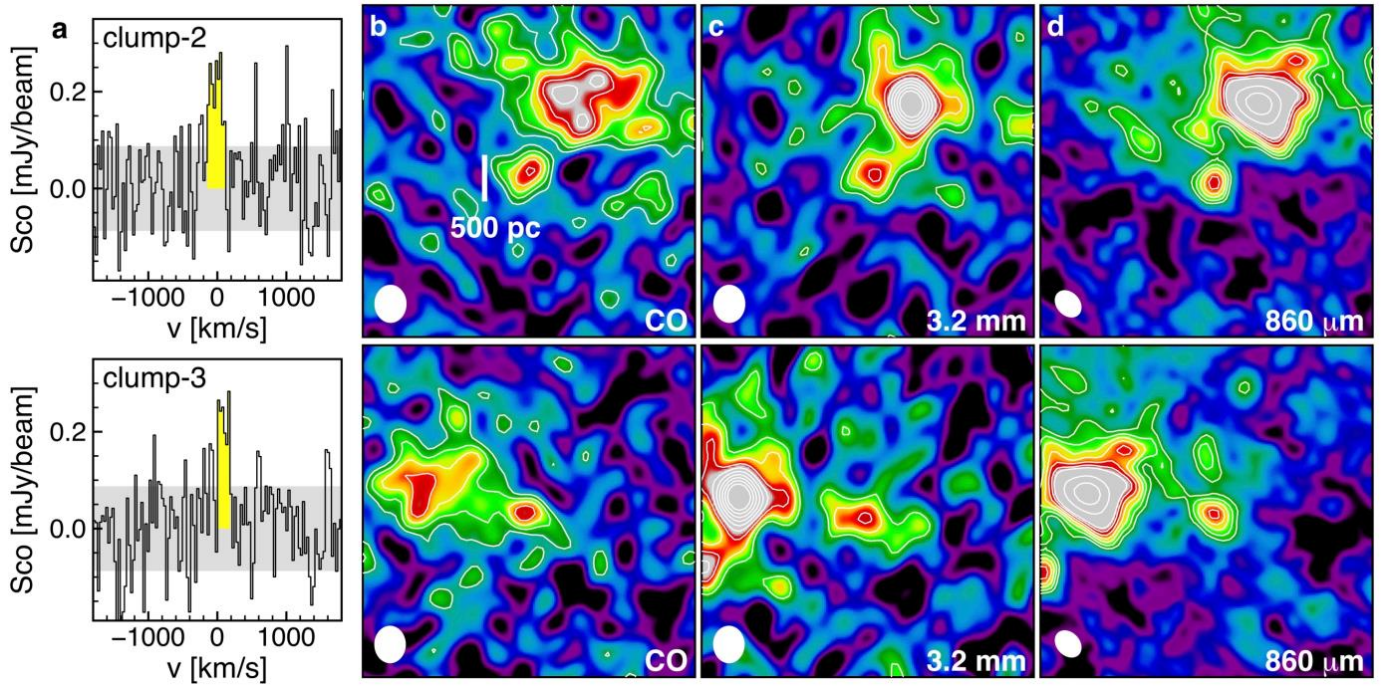

**Figure 2 | Spectra and maps in two clumps. a,** For clump-2 and clump-3, CO spectra are extracted from the briggs-weighted cube with the angular resolution of 0.069″×0.058″. **b, c, d** ALMA maps of CO line, 3.2 mm, 860 μm continuum emission. The CO flux densities are integrated in the velocity range shown by yellow shaded regions of the CO spectra. White filled circles indicate the angular resolution of each map. The contours are plotted at every 1σ from 2σ in CO and 3.2 mm continuum maps and from 4σ in 860 μm continuum map, and at every 5σ from 10σ.

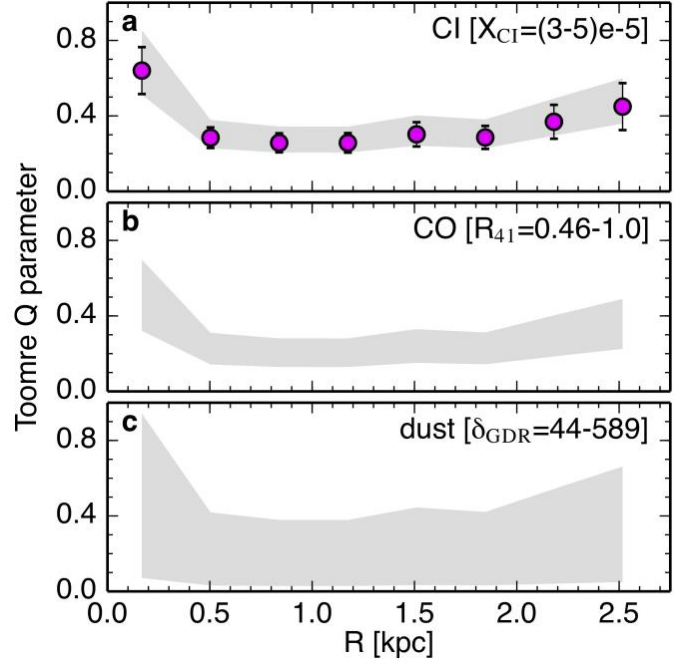

**Figure 3 | Radially-averaged Toomre Q parameter. a,** Magenta circles show the Q values pinned down by the CI-based gas mass. We adopt a carbon abundance relative to molecular hydrogen of $X_{CI}=4\times10^{-5}$. The error bars include the uncertainties of the CI excitation temperature, CO flux measurements, and the kinematic parameters. Gray shaded region indicates to the impact of different carbon abundances, $X_{CI}=(3-5)\times10^{-5}$. **b,c,** We also compute the Q values using the CO-based and dust-based gas mass. Gray shaded regions correspond to the results in CO excitation of $R_{41}=0.46-1.0$ and those in dust-to-gas mass ratio of $\delta_{GDR}=44-589$ (see Methods).

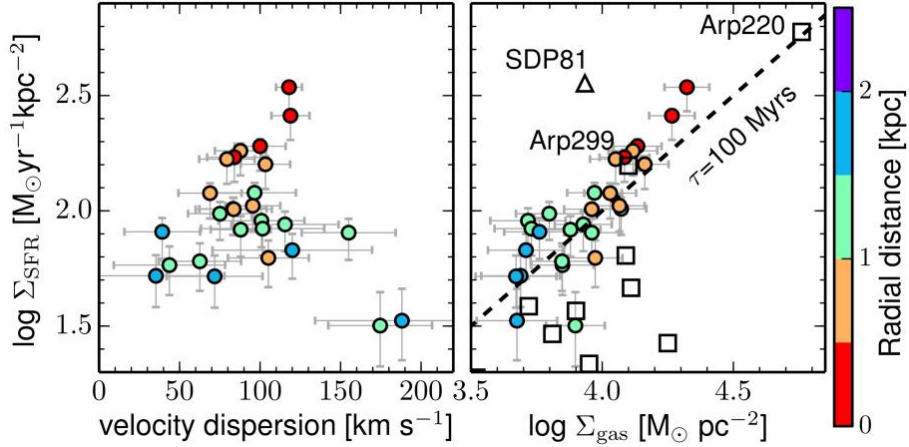

**Figure 4 | Pixel-to-pixel variations of the physical properties in the central 2 kpc region.** Each local velocity dispersion, gas mass surface density, and SFR surface density are computed in a beam area of 0.344 kpc². The color coding indicates a radial distance from the galaxy center. **a,** The local velocity dispersion measured in the central 0.5 kpc region is adversely affected by a beam smearing. **b,** Open squares and open circle show the galaxy-averaged values in nearby starburst galaxies[5] and the resolved value in a gravitationally-lensed galaxy at z=3, SDP 81[16].

**Acknowledgments**
This letter makes use of the following ALMA data: ADS/JAO.ALMA#2017.1.00300.S and 2017.A.00032.S. ALMA is a partnership of ESO (representing its member states), NSF (USA) and NINS (Japan), together with NRC (Canada), NSC and ASIAA (Taiwan), and KASI (Republic of Korea), in cooperation with the Republic of Chile. The Joint ALMA Observatory is operated by ESO, AUI/NRAO and NAOJ. This work was supported by JSPS KAKENHI JP17J04449. We thank J. Baba for helpful discussions about a gravitational instability in SMGs. Data analysis was in part carried out on the common use data analysis computer system at the Astronomy Data Center, ADC, of the National Astronomical Observatory of Japan.

**Author Contributions**
K.T. led the project and reduced the ALMA data. K.T. and D. I. wrote the manuscript. M.S.Y. reduced the LMT data. Other authors contributed to the interpretation and commented on the ALMA proposal and the paper.

**Author Information**
Reprints and permissions information is available at www.nature.com/reprints. The authors declare no competing financial interests. Readers are welcome to comment on the online version of the paper. Correspondence and requests for materials should be addressed to K.T. (tadaki.ken@nao.ac.jp).

**Reference**
1. Swinbank, A. M. et al. Intense star formation within resolved compact regions in a galaxy at z = 2.3. Nature 464, 733-736 (2010).
2. Ikarashi, S. et al. Compact starbursts in z ~ 3-6 submillimeter galaxies revealed by ALMA. Astrophys. J. 810, 133 (2015).
3. Simpson, J. M. et al. The SCUBA-2 Cosmology Legacy Survey: ALMA resolves the rest-frame far-infrared emission of sub-millimeter galaxies. Astrophys. J. 799, 81 (2015).
4. van Dokkum, P. et al. Forming compact massive galaxies. Astrophys. J. 813, 23 (2015).
5. Kennicutt, R. C. Jr. The Global Schmidt law in star-forming galaxies. Astrophys. J. 498, 541-552 (1998).
6. Hughes, D. H. et al. High-redshift star formation in the Hubble Deep Field revealed by a submillimetre-wavelength survey. Nature 394, 241-247 (1998).
7. Barger, A. J. et al. Submillimetre-wavelength detection of dusty star-forming galaxies at high redshift. Nature 394, 248-251 (1998).
8. Chapman, S. C. et al. A redshift survey of the submillimeter galaxy population. Astrophys. J. 622, 772-796 (2005).
9. Bothwell, M. S. et al. A survey of molecular gas in luminous sub-millimetre galaxies. Mon. Not. R. Astron. Soc. 429, 3047-3067 (2013).
10. Ivison, R. J. et al. Herschel-ATLAS: a binary HyLIRG pinpointing a cluster of starbursting protoellipticals. Astrophys. J. 772, 137 (2013).
11. Tacconi, L. J. et al. Submillimeter galaxies at z ~ 2: evidence for major mergers and constraints on lifetimes, IMF, and CO-H2 conversion factor. Astrophys. J. 680, 246-262 (2008).
12. Hodge, J. A. et al. Evidence for a clumpy, rotating gas disk in a submillimeter galaxy at z = 4. Astrophys. J. 760, 11 (2012).
13. Iono, D. et al. Clumpy and extended starbursts in the brightest unlensed submillimeter galaxies. Astrophys. J. 829, L10 (2016).
14. Tadaki, K.-i. et al. Bulge-forming galaxies with an extended rotating disk at z ~ 2. Astrophys. J. 834, 135 (2017).

15. Swinbank, A. M. et al. ALMA resolves the properties of star-forming regions in a dense gas disk at z ~ 3. Astrophys. J. 806, L17 (2015).
16. Sharda, P. et al. Testing Star Formation Laws in a Starburst Galaxy At Redshift 3 Resolved with ALMA. Mon. Not. R. Astron. Soc. 477, 4380-4390 (2018).
17. Bolatto, A. D. et al. The resolved properties of extragalactic giant molecular clouds. Astrophys. J. 686, 948-965 (2008).
18. Cappellari, M. Structure and kinematics of early-type galaxies from integral field spectroscopy. Annu. Rev. Astron. Astrophys. 54, 597-665 (2016).
19. Veale, M. et al. The MASSIVE Survey - V. Spatially resolved stellar angular momentum, velocity dispersion, and higher moments of the 41 most massive local early-type galaxies. Mon. Not. R. Astron. Soc. 464, 356-384 (2017)
20. Naab, T. et al. The ATLAS3D project-XXV. Two-dimensional kinematic analysis of simulated galaxies and the cosmological origin of fast and slow rotators. Mon. Not. R. Astron. Soc. 444, 3357-3387 (2014).
21. Genzel, R. et al. The Sins survey of z ~ 2 galaxy kinematics: properties of the giant star-forming clumps. Astrophys. J. 733, 101 (2011).
22. Bournaud, F. et al. The long lives of giant clumps and the birth of outflows in gas-rich galaxies at high-redshift. Astrophys. J. 780, 57-75 (2014).
23. Mandelker, N. et al. The population of giant clumps in simulated high-z galaxies: in situ and ex situ migration and survival. Mon. Not. R. Astron. Soc. 443, 3675-3702 (2014).
24. Genzel, R. et al. The SINS/zC-SINF survey of z ~ 2 galaxy kinematics: evidence for gravitational quenching. Astrophys. J. 785, 75 (2014).
25. Thompson, T. et al. Radiation pressure-supported starburst disks and active galactic nucleus fueling. Astrophys. J. 630, 167-185 (2005).
26. Cacciato, M. et al. Evolution of violent gravitational disc instability in galaxies: late stabilization by transition from gas to stellar dominance. Mon. Not. R. Astron. Soc. 421, 818-831 (2012).
27. Tacconi, L. J. et al. Phibss: molecular gas content and scaling relations in z ~ 1-3 massive, main-sequence star-forming galaxies. Astrophys. J. 768, 74 (2013).
28. Narayanan, D. et al. The star-forming molecular gas in high-redshift submillimetre galaxies. Mon. Not. R. Astron. Soc. 400, 1919-1935 (2009).
29. Ueda, J. et al. Cold molecular gas in merger remnants. I. Formation of molecular gas disks. Astrophys. J. Suppl. Ser. 214, 1 (2014)
30. Dekel, A. et al. Cold streams in early massive hot haloes as the main mode of galaxy formation. Nature 457, 451-454 (2009).

# Methods

**Sample.** COSMOS AzTEC-1 (AzTEC-1 hereafter) was first discovered as one of the brightest sources in 1.1 mm continuum survey with the bolometer camera AzTEC on the James Clerk Maxwell Telescope (JCMT)[31]. Follow-up observations with Redshift Search Receiver on the Large Millimeter Telescope (LMT) detected CO (4-3) and CO (5-4) lines and determined a spectroscopic redshift of z=4.342, which is also confirmed by the detection of [CII] line in Submillimeter Array observations[32]. Our previous Atacama Large Millimeter/submillimeter Array (ALMA) observations of 860 μm continuum emission at 0.02″ resolution revealed that AzTEC-1 is composed of a compact core, an extended disk and multiple 200 pc clumps on the disk[13]. The half-light radius in the 860 μm emission is $R_{1/2}$=1.1±0.1 kpc. The rest-frame UV continuum emission is not spatially-resolved even with *Hubble Space Telescope (HST)*/WFC3 imaging, suggesting a compact emission with $R_{1/2}$<2.6 kpc as well[33]. Here, a Chabrier initial mass function[34] and cosmological parameters of $H_0$ = 70 km s$^{-1}$ Mpc$^{-1}$, $\Omega_M$=0.3, and $\Omega_\Lambda$=0.7 are assumed. An angular scale of 0.1″ corresponds a physical scale of 670 pc.

**Observations.** In AzTEC-1, we carried out observations of the CO (4-3) emission line at the rest-frame frequency of 461.040 GHz (86.309 GHz in the observed frame) with ALMA band-3 receivers covering the frequency range of 85.4-89.1 GHz and 97.5-101.2 GHz in two array configurations with baseline lengths of 41 m-16.2 km. The shortest 5th percentile baseline is 600 m, corresponding to the maximum recoverable scale of 1.15″ at 86.3 GHz. The observations were executed in 2017 October (C43-10 array configuration) and November (C43-8). On-source time is 5.8 hours and 1.2 hours, respectively. The total observing time including calibration and overhead is 14 hours. We utilize the Common Astronomy Software Application package (CASA)[35] for the data calibration. We first estimated the continuum flux density in the frequency range excluding 86.1-86.5 GHz, and then subtracted it from the visibility data in the *u-v* plane using CASA/*uvcontsub* task. We used CASA/*tclean* task with natural weighting to make a cube with a velocity width of 30 km s$^{-1}$. The resultant angular resolution and the noise level are 0.093″×0.072″ (624 pc×483 pc in physical scale) and 1σ=78 μJy per 30 km s$^{-1}$, respectively. We cleaned down to the 2σ noise level in a circular mask with a radius of 0.4″. We also made s high-resolution cube of CO (4-3) line with a briggs weighting of *robust*=0.5. The angular resolution is 0.069″×0.058″ (470 pc×390 pc) and the noise level is 1σ=87 μJy per 30 km s$^{-1}$. To show the significance of clump detection in AzTEC-1, we also made a 0.055″×0.042″ map of 860 μm continuum emission using the previous ALMA data[13]. We adopted a uv-taper of 0.03″, resulting in the rms level of 47 μJy. We use these high-resolution cube and map only for studying clump properties in Figure 2.

In Extended Data Fig. 1, we show the galaxy-integrated CO (4-3) spectrum extracted within an aperture of 0.8″. The Gaussian fitting gives a line width of FWHM=305±17 km s$^{-1}$. We made the CO moment maps of velocity-integrated intensity, velocity field and velocity dispersion in the velocity range from −315 km s$^{-1}$ to +315 km s$^{-1}$ using CASA/*immoments* task. The 2σ threshold was adopted to exclude noise for creating the velocity field and velocity dispersion maps. We measured a total CO flux of $S_{CO,total}dv$=1.84±0.17 Jy km s$^{-1}$ with a 0.8″ aperture photometry in the velocity-integrated intensity map. The flux measurement error is derived by computing a standard deviation of 300 random apertures in the map. In the LMT observations, the measured flux is $S_{CO}dv$=1.75±0.24 Jy km s$^{-1}$ in the velocity width of $dv$=380 km s$^{-1}$ [32], which is comparable with an ALMA flux of $S_{CO}dv$=1.60±0.13 Jy km s$^{-1}$ in $dv$ =390 km s$^{-1}$.

We also created two 3.2 mm continuum maps with the same angular resolution as the CO (4-3) cubes by excluding the CO frequency range. The rms level is 3.0 μJy beam$^{-1}$ in the 0.093″×0.072″ map and 3.3 μJy beam$^{-1}$ in the 0.069″×0.058″ map. We derived a total flux density of $S_{3.2mm}$=273±41 μJy with an aperture of 0.8″, which is consistent with a 3 mm continuum flux density of $S_{3mm}$=300±40 μJy from Plateau de Bure interferometer observations with 6″ beam[36].

We made follow-up observations of CI (1-0) line at 92.134 GHz in the observed frame and CI (2-1) line at 151.511 GHz with ALMA band-3 and Band-4 receivers in 2018 March. We reduced the data in a similar way as the CO (4-3) data and created a 30 km s$^{-1}$ cube and 2.1 mm continuum map with briggs weighting (robust parameter of +0.5). The angular resolution is 1.7″×1.1″ in the CI (1-0) map and 0.8″×0.7″ in the CI (2-1) map. The noise level is 1σ=0.49 mJy per 30 km s$^{-1}$ in the CI (1-0) cube, 1σ=0.38 mJy per 30 km s$^{-1}$ in the CI (2-1) cube and 1σ=20 μJy in the 2.1 mm continuum map. For flux measurements of both CI lines, we integrated the cubes in the same velocity range as the CO (4-3) and obtained the peak fluxes. We also made a natural weighted CI (2-1) map with the same angular resolution as the CI (1-0) map using CASA/*imsmooth* task, which is used for obtaining a CI (1-0)/CI (2-1) line ratio. In Extended Data Figure 1 and Table 1, we show the line spectra and tabulate the measured line fluxes and luminosities. The 2.1 mm continuum flux density is $S_{2.1mm}$=989±20 μJy. We also detected the CO (7-6) emission line at 151.007 GHz, which is located near CI (2-1) line, but do not use this information.

**Global spectral energy distribution (SED) properties of AzTEC-1.** We collected the photometric data for AzTEC-1 from the latest multi-wavelength catalogs (Subaru[37], VISTA[37], Spitzer[37], Hershel[38,39], VLA[40]). After excluding marginal detections below 5σ and adding our ALMA photometry at 860 μm, 2.1mm and 3.2 mm, we successfully constrained the global SED from optical to radio (Extended Data Fig. 2). Using the *MAGPHYS* code[41,42], we fit the observed SED to stellar population synthesis models[43] with taking into account dust attenuation and dust emission in a physically consistent way. The best-fit SED model indicates AzTEC-1 is a massive, star-forming galaxy with a stellar mass of $M_{star}$=(9.7$^{+0.0}_{-2.5}$)×10$^{10}$ $M_\odot$ and a star formation rate of SFR=1169$^{+11}_{-274}$ $M_\odot$yr$^{-1}$. The dust emission is characterized by a total infrared luminosity of $L_{dust}$=(1.9$^{+0.0}_{-0.3}$)×10$^{13}$ $L_\odot$, a dust mass of $M_{dust}$=(1.1$^{+0.2}_{-0.1}$)×10$^{9}$ $M_\odot$ and a dust temperature of $T_{dust}$=43$^{+3}_{-2}$ K. The uncertainties are based on the 2.5th-97.5th percentile range of the probability distributions.

**Gas mass.** For gas mass estimates based on CO (4-3) luminosity, there are uncertainties about a CO excitation, $R_{41}$=$L'_{CO(4-3)}/L'_{CO(1-0)}$, and a CO-to-H$_2$ conversion factor, $\alpha_{CO}$=$M_{gas}/L'_{CO}$. Alternatively, [CI] line is used as an excellent optically thin tracer of cold molecular gas in nearby and high-redshift galaxies[44-48]. The measured [CI] line luminosity pins down the total molecular gas mass in AzTEC-1. First, we estimated an excitation temperature of $T_{ex}$=27.7±4.8 K from the CI (1-0)/CI (2-1) line ratio of $R_{CI}$=0.52±0.13 in the 1.7″×1.1″-resolution maps as $T_{ex}$=38.8 K/ln(2.11/$R_{CI}$) [45]. Then, using the CI (2-1) flux in the 0.8″×0.7″ map, we computed a neutral carbon mass of $M_{CI}$=(1.7±0.3)×10$^{7}$ $M_\odot$ as $M_{CI}$=4.566×10$^{-4}$$Q(T_{ex})$×1/5×exp(62.5/$T_{ex}$)$L'_{CI(2-1)}$ where $Q(T_{ex})$=1+3exp(-23.6/$T_{ex}$)+5exp(-62.5/$T_{ex}$) is the partition function[45]. The uncertainty of a neutral carbon mass includes the flux measurement error and the error of excitation temperature. We eventually obtained a gas mass of $M_{CI,gas}$=(7.2±1.3)×10$^{10}$ $M_\odot$ by adopting a carbon abundance relative to molecular hydrogen of $X_{CI}$=4×10$^{-5}$, which is the average between the typical value of 3×10$^{-5}$ in normal star-forming galaxies [44-48] and the elevated value of 5×10$^{-5}$ in the central region of local starburst galaxy, M82[49]. The gas-to-dust mass ratio is $M_{CI,gas}/M_{dust}$=65±17, which is smaller than the average value of δ$_{GDR}$=120±28 in 18 nearby starburst galaxies[50], but is still in the range between the 5th and the 95th percentile, δ$_{GDR}$=44-589. The CO (4-3) luminosity also gives a gas mass of $M_{CO,gas}$=(6.6±0.6)×10$^{10}$×($\alpha_{CO}$/0.8)×(1.0/$R_{41}$) $M_\odot$. If the CO (4-3) line is thermalized ($R_{41}$=1) and the conversion factor of $\alpha_{CO}$=0.8 $M_\odot$ (K km s$^{-1}$ pc$^2$)$^{-1}$ is used[11,12,51,52], the CO-based gas mass is similar to the CI-based gas mass. For the consistency between CI and CO, we adopt a gas excitation of $R_{41}$=0.91, which is larger than the average value for SMGs at high-redshift ($R_{41}$=0.46) and comparable with the average value for quasi-stellar objects ($R_{41}$=0.87)[53]. The assumption of $\alpha_{CO}$=4 $M_\odot$ (K km s$^{-1}$ pc$^2$)$^{-1}$, commonly used in normal star-forming galaxies[51], is not appropriate since the CO-based gas mass substantially exceeds the CI-based one. A numerical simulation shows that CO in dense clumps is more excited, compared to an entire disk[54]. If the gas is thermalized with $R_{41}=1$, the gas mass of clumps can be 10 percent smaller.

**SFR and Gas mass surface density.** We obtained the total SFR, $SFR_{total}$, and gas mass, $M_{gas,total}$, in AzTEC-1 as mentioned above. Since the 860 µm continuum flux density, $S_{860µm}$, traces star formation, we compute SFR surface densities in each pixel as $\Sigma_{SFR} = SFR_{total} \times (S_{860µm}/S_{860µm,total}) / \Omega_{beam}$ where $S_{860µm,total}=16.9\pm0.7$ mJy and $\Omega_{beam}$ is the effective beam area of 0.344 kpc$^2$. Here, we use the 860 µm continuum map with a pixel scale of 0.07″ to avoid oversampling although the original pixel scale is 0.01″. The uncertainties of $\Sigma_{SFR}$ include both flux measurement errors of $S_{860µm}$ and $S_{860µm,total}$ as well as the error of SED modeling. In a similar way, using the CO (4-3) map, we derive gas mass surface densities as $\Sigma_{gas} = M_{gas,total} \times (S_{CO}dv/S_{CO,total}dv) / \Omega_{beam}$.

**Disk modeling and dynamical mass.** We fit the natural-weighted CO cube with dynamical models of a disk galaxy using GalPaK$^{3D}$ code[55]. We adopt a thick exponential disk with an arctan rotation curve, $v(R) \propto v_{max}\arctan(R/R_t)$ where $v_{max}$ is the maximum circular velocity and $R_t$ is the turnover radius. Model galaxies have 10 free parameters of $x$, $y$ positions, systematic velocity $v_{sys}$, line flux $Sdv$, half-light radius $R_{1/2}$, $R_t$, inclination $i$, position angle PA, $v_{max}$, and velocity dispersion $\sigma_0$. They are convolved with the clean beam and are fitted to the data cube using the Markov Chain Monte Carlo (MCMC) algorithm. In Extended Data Fig. 3, we show the CO spectra extracted along the kinematic major axis in the observed cube together with the best-fit model. The observed CO kinematics is well characterized by a rotating disk model. The best-fitting values are $Sdv=1.88^{+0.02}_{-0.01}$ Jy km s$^{-1}$, $R_{1/2}=1.05^{+0.02}_{-0.02}$ kpc, $R_t=0.18^{+0.03}_{-0.03}$ kpc, $i=44^{+1}_{-1}$ degrees, PA$=-64^{+1}_{-1}$ degrees, $v_{max}=227^{+5}_{-6}$ km s$^{-1}$ and $\sigma_0=74^{+1}_{-1}$ km s$^{-1}$. We adopted the median and the 95% confidence interval of the last 60% of the Malkov Chain Monte Carlo (MCMC) chain for 20000 iterations as the best-fit values and the uncertainties (Extended Data Figure 4). For symmetric oblate disks, the inclination corresponds to the projected minor-to-major-axis ratio of $q_{obs}=0.73$ as $\sin^2(i)=(1-q_{obs}^2)/(1-q_{int}^2)$, if we assume a disk thickness of $q_{int}=0.15$.

**Toomre Q parameter.** In a thin rotating gas disk with epicyclic frequency of $\kappa$, the dispersion relation for axisymmetric perturbations is given by $\omega^2=\kappa^2-2\pi G\Sigma_{gas}|k|+\sigma_{0,gas}^2 k^2$ where $\omega$ is the growth rate, $k$ is the wavenumber of the perturbation[56-58]. The perturbations exponentially grow in time when $\omega^2<0$, leading to gravitational collapse of gas clouds. This condition is practically estimated by comparing the measured Toomre Q parameter of $Q=\kappa\sigma_{0,gas}/(\pi G\Sigma_{gas})$ to the threshold of $Q_{cri}=1$ for a thin gas disk or $Q_{cri}=0.67$ for a thick disk[21,26]. When the disk consists of two components (gas and stars) with the same velocity dispersion, the threshold value increases up to $Q_{cri,2com}=1.3$ [59]. The self-gravity of gas overcomes the repelling forces by pressure and differential rotation when $Q<Q_{cri}$. Exploiting the maximum circular velocity and the ALMA maps of molecular gas mass surface density and velocity dispersion without correction for beam-smearing, we estimated local $Q$ parameters in each pixel assuming a flat rotation curve with $\kappa=1.4v_{max}/R$[58]. In Figure 3, we computed radially-averaged CO fluxes along elliptical rings with the axis ratio of 0.73.

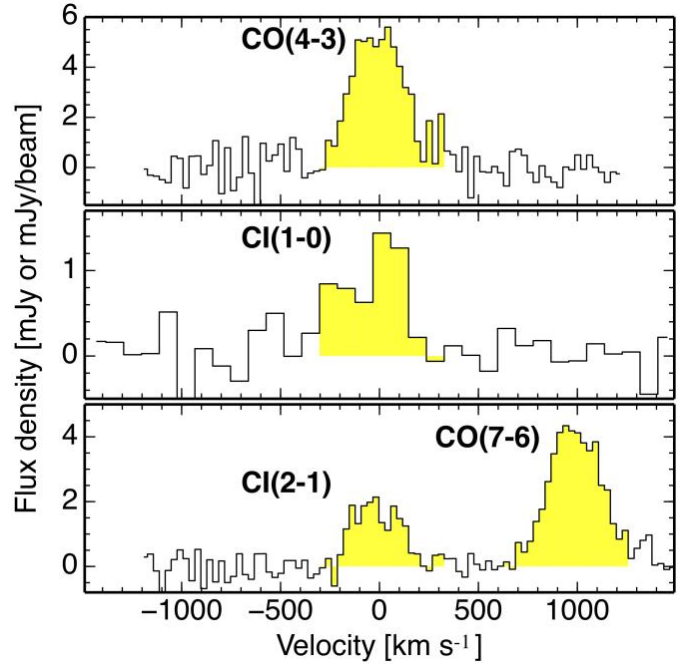

**Extended Data Figure 1 | A galaxy-integrated CO (4-3), CO (1-0), CI(2-1) spectra in AzTEC-1.** The CO (4-3) spectrum is extracted within an aperture of 0.8″ in the natural weighted map. The CI (1-0) and CI(2-1) spectra are extracted from the peak positions in the 1.7″×1.1″ resolution map and 0.8″×0.7″-resolution one, respectively. Yellow shaded regions show the velocity range of $v=\pm 315$ km s$^{-1}$ where the velocity-integrated fluxes are measured.

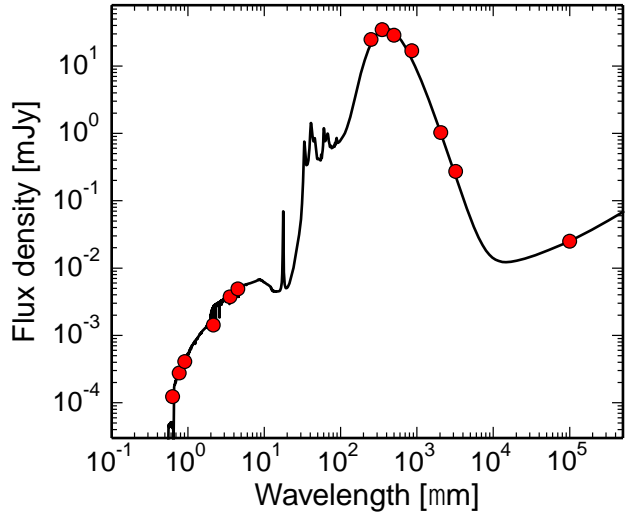

**Extended Data Figure 2 | A galaxy-integrated SED in AzTEC-1.** Red circles show the photometric data from Subaru ($r'$, $i'$, $z'$)[37], VISTA ($K_s$)[37], Sptizer (3.6 μm, 4.4 μm)[37], Herschel (250 μm, 350 μm, 500 μm)[38,39], ALMA (860 μm, 2.1 mm, 3.2 mm), JVLA (10 cm)[40]. A black line indicates the best-fit SED model from *MAGPHYS*[41,42].

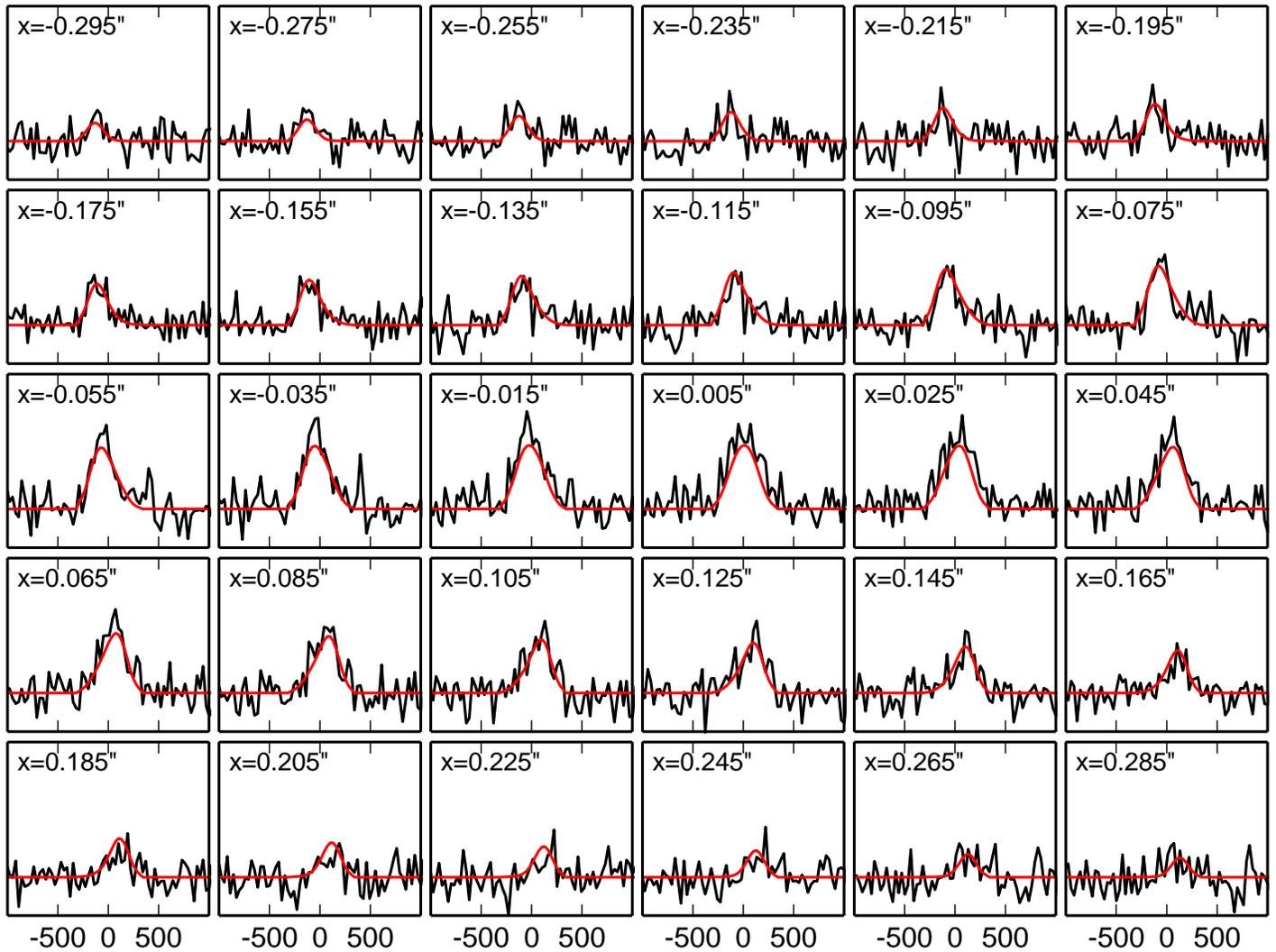

**Extended Data Figure 3 | CO spectra along the kinematics major axis.** Spectra are extracted at the position angle of PA=-64 degrees. The spatial offset from the galaxy center is shown at the upper left of each panel. Red lines indicate the spectra of the best-fit dynamical model produced by GalPaK$^{3D}$.

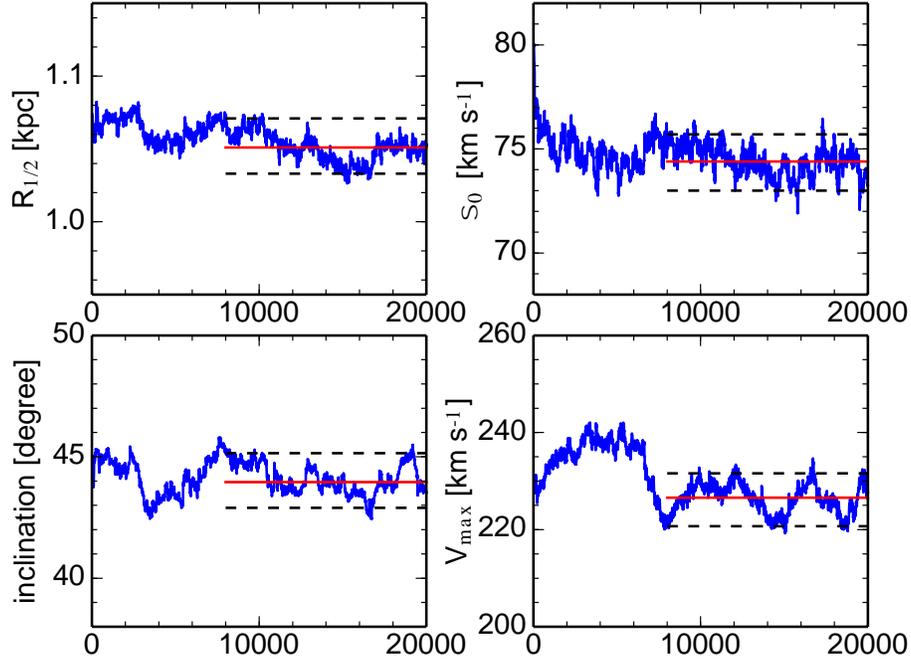

**Extended Data Figure 4 | Full MCMC chain for 20,000 iterations.** Red solid line and black dashed lines indicate the median and the 95% confidence interval of the last 60% of the MCMC chain.

**Extended Data Table 1 | Line fluxes in AzTEC-1.**

| Line | Frequency (GHz) | $S_{CO}dv$ (Jy km s$^{-1}$) | $L'_{line}$ ($10^{10}$K km s$^{-1}$ pc$^2$) |
|---|---|---|---|
| CO (4-3) | 461.041 | 1.84±0.17* | 8.21±0.78 |
| CI (1-0) | 492.161 | 0.45±0.08† | 1.76±0.30 |
| CI (2-1) | 809.342 | 0.49±0.09‡ | 0.70±0.13 |
| CI (2-1) | 809.342 | 0.63±0.11† | 0.92±0.16 |

* The flux within a 0.8″ aperture in the 0.093″×0.072″ map.
† The peak flux in the 1.7″×1.1″ map.
‡ The peak flux in the 0.8″×0.7″ map.

31. Scott, K. B. et al. AzTEC millimetre survey of the COSMOS field – I. Data reduction and source catalogue. Mon. Not. R. Astron. Soc. 385, 12225-2238 (2008).
32. Yun, M. S. et al. Early science with the Large Millimeter Telescope: CO and [C II] emission in the z=4.3 AzTEC J095942.9+022938 (COSMOS AzTEC-1). Mon. Not. R. Astron. Soc. 454, 3485-3499 (2015).
33. Toft, S. et al. Submillimeter galaxies as progenitors of compact quiescent galaxies. Astrophys. J. 782, 68 (2014).
34. Chabrier, G. The galactic disk mass function: reconciliation of the Hubble Space Telescope and nearby determinations. Astrophys. J. 586, L133-L136 (2003).
35. McMullin, J. P., Waters, B., Schiebel, D., Young, W. & Golap, K. CASA architecture and applications. ASP Conf. Ser. 376, 127-130 (2007).
36. Smolcic V. et al. The Redshift and Nature of AzTEC/COSMOS 1: A Starburst Galaxy at z = 4.6. Astrophys. J. 731, L27 (2011).
37. Laigle, C. et al. The COSMOS2015 catalog: exploring the 1<z<6 universe with half a million galaxies. Astrophys. J. Suppl. S. 224, 24 (2016).
38. Roseboom, I. G. et al. The Herschel Multi-tiered Extragalactic Survey: SPIRE-mm photometric redshifts. Mon. Not. R. Astron. Soc. 419, 2758-2773 (2012).
39. Oliver, S. J. et al. The Herschel Multi-tiered Extragalactic Survey: HerMES. Mon. Not. R. Astron. Soc. 424, 1614-1635 (2012).
40. Smolcic, V. et al. The VLA-COSMOS 3 GHz Large Project: Continuum data and source catalog release. Astron. Astrophys. 602, A1 (2017).
41. da Cunha, E., Charlot, S. & Elbaz, D. A simple model to interpret the ultraviolet, optical and infrared emission from galaxies. Mon. Not. R. Astron. Soc. 388, 1595-1617 (2008).
42. da Cunha, E. et al. An ALMA survey of sub-millimeter galaxies in the Extended Chandra Deep Field South: physical properties derived from ultraviolet-to-radio modeling. Astrophys. J. 806, 110 (2015).
43. Bruzual, G. & Charlot, S. Stellar population synthesis at the resolution of 2003. Mon. Not. R. Astron. Soc. 344, 1000-1028 (2003).
44. Papadopoulos, P. P., Thi, W.-F., Viti. S. CI lines as tracers of molecular gas, and their prospects at high redshifts. Mon. Not. R. Astron. Soc. 351, 147-160 (2004).
45. Weiss, A. et al. Gas and dust in the Cloverleaf quasar at redshift 2.5. Astron. Astrophys. 409, L41-L45 (2003).
46. Weiss, A. et al. Atomic carbon at redshift ~2.5. Astron. Astrophys. 429, L25-L28 (2005).
47. Danielson, A. L. R. et al. The properties of the interstellar medium within a star-forming galaxy at z = 2.3. Mon. Not. R. Astron. Soc. 410, 1687-1702 (2011).
48. Bothwell, M. S. et al. ALMA observations of atomic carbon in z ~ 4 dusty star-forming galaxies. Mon. Not. R. Astron. Soc. 466, 2825-2841 (2017).
49. White, G. J. et al. CO and CI maps of the starburst galaxy M 82. Astron. Astrophys. 284, L23-26 (1994).
50. Wilson, C. et al. Luminous infrared galaxies with the submillimeter array. I. Survey overview and the central gas to dust ratio. Astrophys. J. Suppl. Ser. 178, 189-224 (2008).
51. Bolatto, A. D., Wolfire, M. & Leroy, A. K. The CO-to-H2 conversion factor. Annu. Rev. Astron. Astrophys. 51, 207-268 (2013).
52. Downes, D. & Solomon, P. M. Rotating nuclear rings and extreme starbursts in ultraluminous galaxies. Astrophys. J. 507, 615-654 (1998).
53. Carilli, C. L. & Walter, F. Cool gas in high-redshift galaxies. Annu. Rev. Astron. Astrophys. 51, 105-161 (2013).
54. Bournaud, F. et al. Modeling CO emission from hydrodynamic simulations of nearby spirals, starbursting mergers, and high-redshift galaxies. Astron. Astrophys. 575, A56 (2015).
55. Bouche, N. et al. GalPak3D: A Bayesian Parametric Tool for Extracting Morphokinematics of Galaxies from 3D Data. Astrophys. J. 150, 92 (2015).
56. Toomre, A. On the gravitational stability of a disk of stars. Astrophys. J. 139, 1217-1238 (1964).
57. Wang, B. et al. Gravitational instability and disk star formation. Astrophys. J. 427, 759-769 (1994).
58. Binney, J. & Tremaine, S. Galactic Dynamics 2nd edn (Princeton Univ. Press, 2008).

59. Romeo, A. B. & Wiegert, J. The effective stability parameter for two-component galactic discs: is $Q^{-1} \approx Q^{-1}_{stars} + Q^{-1}_{gas}$?. Mon. Not. R. Astron. Soc. 416, 1191-1196 (2011).